\documentclass[%
reprint,
superscriptaddress,
amsmath,amssymb,
prapplied,
]{revtex4-1}

\usepackage{graphicx}
\usepackage{color}
\usepackage{dcolumn}
\usepackage{bm}
\usepackage{upgreek}
\usepackage{hyperref}
\usepackage[normalem]{ulem}

\begin{document}

	\title{Fast gate-based readout of silicon quantum dots using Josephson parametric amplification}
	
	\author{S. Schaal}
	\email{simon.schaal.15@ucl.ac.uk}
	\affiliation{London Centre for Nanotechnology, UCL, London WC1H 0AH, United Kingdom}
	
	\author{I. Ahmed}
	\altaffiliation[Present address: ]
	{Dept.\ of Electrical \& Electronic Engineering, Univ.\ of Dhaka, Dhaka 1000, Bangladesh}
	\affiliation{Cavendish Laboratory, Univ.\ of Cambridge, J. J. Thomson Ave., Cambridge, CB3 0HE, U.K.}
	\author{J. A. Haigh}
	\affiliation{Hitachi Cambridge Laboratory, J.J. Thomson Avenue, Cambridge CB3 0HE, United Kingdom}
	
	\author{L. Hutin}
	\author{B. Bertrand}
	\author{S. Barraud}
	\author{M. Vinet}
	\affiliation{CEA, LETI, Minatec Campus, F-38054 Grenoble, France}
	
	\author{C.--M. Lee}
	\author{N. Stelmashenko}
	\author{J. W. A. Robinson}
	\affiliation{Dept.\ of Materials Science \& Metallurgy, Univ.\ of Cambridge, 27 Charles Babbage Road, CB3 0FS, U.K.}
	
	\author{J. Y. Qiu}
	\altaffiliation[Present address: ]
	{Massachusetts Institute of Technology, Cambridge, MA 02139, USA}
	\affiliation{Quantum Nanoelectronics Laboratory, Dept.\ of Physics, 	Univ.\ of California, Berkeley CA 94720, USA}	
	\author{S. Hacohen-Gourgy}
	\author{I. Siddiqi}
	\affiliation{Quantum Nanoelectronics Laboratory, Dept.\ of Physics, 	Univ.\ of California, Berkeley CA 94720, USA}	
	
	\author{M. F. Gonzalez-Zalba}
	\affiliation
	{Hitachi Cambridge Laboratory, J.J. Thomson Avenue, Cambridge CB3 0HE, United Kingdom}
	
	\author{J. J. L. Morton}
	\email{jjl.morton@ucl.ac.uk}
	\affiliation{London Centre for Nanotechnology, UCL, London WC1H 0AH, United Kingdom}
	\affiliation
	{Dept.\ of Electronic \& Electrical Engineering, UCL, London WC1E 7JE, United Kingdom}

	\date{\today}
	
	\begin{abstract}
		Spins in silicon quantum devices are promising candidates for large-scale quantum computing. 		
		Gate-based sensing of spin qubits offers compact and scalable readout with high fidelity, however further improvements in sensitivity are required to meet the fidelity thresholds and measurement timescales needed for the implementation of fast-feedback in error correction protocols. Here, we combine radio-frequency gate-based sensing at 622~MHz with a Josephson parametric amplifier (JPA), that operates in the 500--800~MHz band, to reduce the integration time required to read the state of a silicon double quantum dot formed in a nanowire transistor. Based on our achieved signal-to-noise ratio (SNR), we estimate that singlet-triplet single-shot readout with an average fidelity of 99.7\% could be performed in 1~$\upmu$s, well-below the requirements for fault-tolerant readout and 30 times faster than without the JPA. Additionally, the JPA allows operation at a lower RF power while maintaining identical SNR. We determine a noise temperature of 200~mK with a contribution from the JPA (25\%), cryogenic amplifier (25\%) and the resonator (50\%), showing routes to further increase the read-out speed.  		
	\end{abstract}

	\maketitle

	Quantum computers require high-fidelity qubit measurement, which must be performed on a timescale faster than the decoherence time to perform quantum error correction~\cite{Fowler2012}.
	Spin qubits formed in quantum dots (QDs) or donors in silicon are one of the most promising platforms for scalable quantum information processing due to their long coherence times and large integration density~\cite{Vandersypen2017,Veldhorst2017,Li,Pica2015,Hill2015,Cai2019}. 
	In such devices, readout has been typically achieved using nearby electrometers to detect the spin state via spin-to-charge conversion based on spin dependent tunneling~\cite{Elzerman2004,Morello2010} or Pauli spin blockade~\cite{Prance2012,Barthel2009,Zhao2018a}, with radio-frequency (RF) single-electron transistors being the most sensitive electrometers to date~\cite{Schoelkopf1998,Brenning2006}.    
	When scaling to large arrays of dense qubits~\cite{Jones2018,Veldhorst2017,Vandersypen2017,Li} space for additional electrometers and reservoirs is  limited: Gate-based dispersive RF readout eliminates the need for additional sensor structures~\cite{Colless2013,Gonzalez-Zalba2015,Petersson2010a} and reservoirs (using Pauli spin blockade)~\cite{Schroer2012a,Betz2015,Urdampilleta2015} by embedding the gates that define the quantum dot into a resonant circuit. This gate-based readout relies on detecting a shift in the phase of the RF signal reflected by the resonator which is proportional to the quality factor ($Q_\mathrm{load}$), the square of the gate coupling $\alpha$ and the inverse of the total capacitance of the resonator~\cite{Ahmed2018}. 
	Recently, single-shot readout of the singlet-triplet states in a double QD has been demonstrated with gate-based sensors, using a variety of resonator parameters to achieve a range of readout fidelities (for a given integration time): $73\%$ (2.6~ms)~\cite{West2019}, $82.9\%$ ($300\, \upmu$s)~\cite{Pakkiam2018}, $98\%$ ($6\, \upmu$s)~\cite{Zheng2019} to $99\%$ (1~ms; using ancillary `sensor' QD and reservoir)~\cite{Urdampilleta2018}.
		
	Amplifiers based on Josephson junctions have greatly improved signal-to-noise ratios (SNRs) in the field of superconducting circuits~\cite{Slichter2016,Vijay2012,Vijay2011,Eichler2012,Chen2012a,Hatridge2011} --- they typically operate at frequencies of several GHz and near the quantum limit of noise introduced by the amplifier (or indeed below, for a single quadrature using squeezing)~\cite{Caves2012,Castellanos-Beltran2008,Castellanos-Beltran2007,Eichler2011,Bergeal2010,Macklin2015}.
	Adopting such approaches in the measurement of QDs at RF/microwave frequencies is expected to lead to corresponding improvements in SNR. While this can in principle be achieved at operating frequencies of 4--8~GHz that are typical for Josephson-junction based amplifiers, as demonstrated using an InAs double QD, Josephson parametric amplifier (JPA) and coplanar waveguide resonator~\cite{Stehlik2015}, lower frequency operation ($\lesssim1$~GHz) becomes necessary~\footnote{operating at a frequency $f>\Delta_\mathrm{c}/4h$ comparable to the tunnel coupling $\Delta_\mathrm{c}$ results in back-action of the resonator onto the quantum dot device in form of fast voltage oscillations that manifest as Landau-Zener transitions} for studying lower QD tunneling rates, at which exchange interaction is more easily controlled, and for enabling off-chip resonator fabrication.
	Suitable amplifiers are available in such a frequency range, for example: a JPA operating at 600~MHz with a noise temperature of $T_\mathrm{JPA}=105$~mK~\cite{Simbierowicz2018} or a SQUID amplifier chain with $T_\mathrm{SQUID}=52$~mK at 538~MHz~\cite{Muck2001}. Building on such developments, readout of a GaAs based quantum dot at 196~MHz with a noise temperature of 490~mK was recently reported using a SQUID amplifier~\cite{Schupp2018}.
	
	In this Letter, we combine RF capacitive gate-based sensing of silicon QDs with Josephson parametric amplification to push the bounds of SNR that can be achieved using this technique. We use a well-matched lumped-element high $Q$ resonator containing a NbN spiral inductor and a JPA that operates in the 500--800~MHz band and obtain an overall noise temperature $T_\mathrm{noise}=200$~mK at 621.9~MHz. 
	We benchmark the sensitivity of the method using electronic transitions in a silicon multi-dot device with large gate-coupling (wrap-around geometry) fabricated following CMOS processes.
	When using the JPA at an inter-dot charge transition, we find an improvement of a factor of 7 in the SNR and a minimum integration time of 80~ns. Based on our measurements, we estimate an average single-shot fidelity of $99.7\%$ would be possible in $\tau_\mathrm{int}=32\, \upmu$s without the JPA, yet only $\tau_\mathrm{int}=1\, \upmu$s with the JPA. These improvements in readout speed enable implementations of error correction codes and fast feedback for silicon-based quantum devices.
	
	
	\begin{figure}
		\centering
		\includegraphics[width=\linewidth]{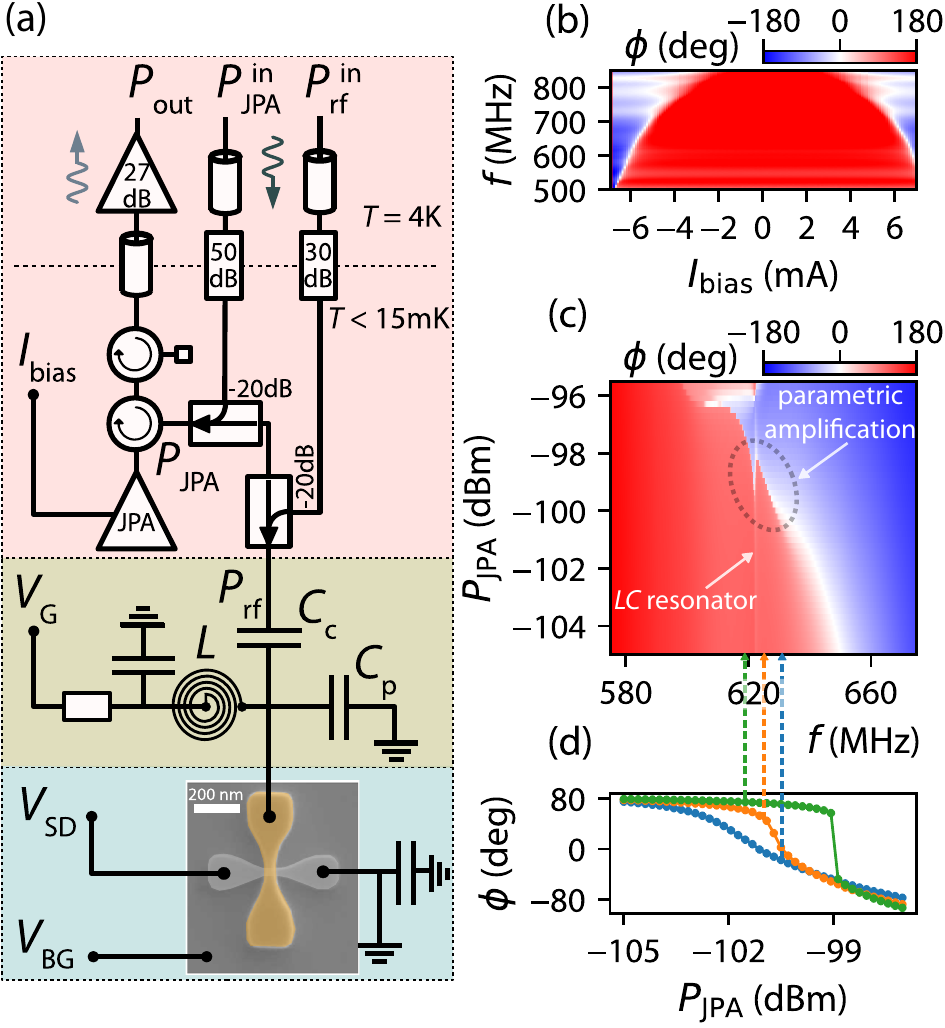}
		\caption{{\bf Set-up and Josephson Parametric Amplifier (JPA)}. \textbf{(a)} Schematic of the quantum dot readout setup consisting of microwave components (including a JPA), readout resonator and CMOS quantum dot device (false-colored SEM micro-graph). \textbf{(b)} Phase response of the JPA as a function of flux bias $I_\mathrm{bias}$ demonstrating frequency tuning of the JPA. \textbf{(c)} JPA phase response as a function pump power and frequency. The regime of parametric amplification and the readout resonator are highlighted and are found just below the power at which the JPA enters the bi-stable regime. \textbf{(d)} JPA transfer function obtained from line-cuts at indicated frequencies in (c). The bi-stable regime is observed as an abrupt jump in $\phi$.}
		\label{fig:jpa}
	\end{figure}
	
	A schematic of how the JPA is embedded into the cryogenic reflectometry setup is shown in Fig.~\ref{fig:jpa}(a). The setup consists of i) the cryogenic RF delivery and amplification chain including the JPA (pink background); ii) a lumped-element $LC$ resonator (green); and iii) the silicon quantum dot device (blue), see~\cite{SuppMat} for details.
	In gate-based sensing, the device is embedded in a $LC$ resonator which is probed using an RF tone with power $P_\mathrm{rf}$ at resonant frequency $f_\mathrm{rf}$. 
	At this frequency, changes in device capacitance, $\Delta C_\mathrm{d}$, due to cyclic single-electron tunneling produce changes in the reflection coefficient $\Delta \Gamma = |\frac{\partial \Gamma}{\partial C_\mathrm{d}}\Delta C_\mathrm{d}|$~\cite{Mizuta2016}. This effect translates into a change in the reflected power with an SNR given by
	\begin{align*}
	\mathrm{SNR}=|\Delta\Gamma|^2 \frac{P_\mathrm{rf}}{P_\mathrm{n}},
	\end{align*}
	where $P_\mathrm{n}$ is the noise power. 
	The optimal SNR is achieved by maximizing $\Delta\Gamma$ (i.e.\ large loaded quality factor and small parasitic capacitance, which translates into a large resonator impedance, combined with large gate coupling and a well matched resonator)~\cite{Ahmed2018}, maximizing $P_\mathrm{rf}$ (while remaining below power broadening) and minimizing $P_\mathrm{n}$.
	
	The noise power for an amplifier with gain $G$ can be defined as $P_\mathrm{n, out}=G k_B (T_\mathrm{sys}+T_\mathrm{n})B$ where $T_\mathrm{sys}$ and $T_\mathrm{n}$ are the system and amplifier noise temperature (noise added by the amplifier) respectively, $k_B$ is Boltzmann's constant and $B$ is the amplifier bandwidth.
	In semiconductor QD measurements, cryogenic high electron mobility transistor (HEMT) amplifiers operating at 4~K typically limit the effective noise temperature ($T_\mathrm{HEMT}$) to a few Kelvin. By including an additional amplifier (such as a JPA) with gain $G_\mathrm{JPA} (\gg 1)$ and lower noise temperature ($T_\mathrm{JPA}$) at the beginning of the amplification chain, the effective noise temperature $T_\mathrm{noise}$ can be reduced:
	\begin{align}
	T_\mathrm{noise}=T_\mathrm{sys} + T_\mathrm{JPA}+\frac{T_\mathrm{HEMT}}{G_\mathrm{JPA}}. 
	\label{eq:Tsys}
	\end{align}
	For a JPA operating at $T=10\, $mK 
	we expect a minimum of $T_\mathrm{JPA}=\frac{\hbar \omega}{2 k_B} \coth\left(\frac{\hbar \omega}{2k_B T}\right)=16.5\, $mK.
	
	As shown in Fig.~\ref{fig:jpa}(a), the RF signal reflected from the quantum device passes the JPA (which works in reflection) via a circulator and is further amplified at $4\, $K followed by additional amplification and quadrature demodulation for measurement at room temperature (not shown).
	Our JPA is a low quality factor ($Q_\mathrm{JPA}< 100$) superconducting resonator consisting of a SQUID loop array with tunable inductance shunted by a fixed capacitance~\cite{Vijay2009}, and is tunable in frequency from 500--800~MHz, as shown in Fig.~\ref{fig:jpa}(b), by passing a current $I_\mathrm{bias}$ through a coil that changes the flux through the nearby SQUIDs.
	The JPA is pumped via the signal input port and with power $P_\mathrm{JPA}$, delivered using a separate microwave line. At low drive power the JPA behaves like a linear resonator, while at high power the non-linearity of the Josephson junctions manifests in a frequency shift of the JPA to lower frequency until eventually the JPA reaches a bi-stable regime~\cite{Vijay2009}. The JPA phase response as a function of pump power and frequency is shown in Fig.~\ref{fig:jpa}(c) where the additional phase shift originating from the $LC$ resonator and the regime useful for parametric amplification are indicated.
	In this regime, the phase of the reflected pump signal varies rapidly with the pump power, as shown in Fig.~\ref{fig:jpa}(d), which represents the transfer function of the JPA. When biased at this point, changes in the pump amplitude due to a small signal lead to large changes in the reflected phase, such that the gain is determined by the gradient of the transfer function while the width in power sets the dynamic range. For small modulations of $P_\mathrm{JPA}$, the response is linear.

	\begin{figure}
		\centering
		\includegraphics[width=\linewidth]{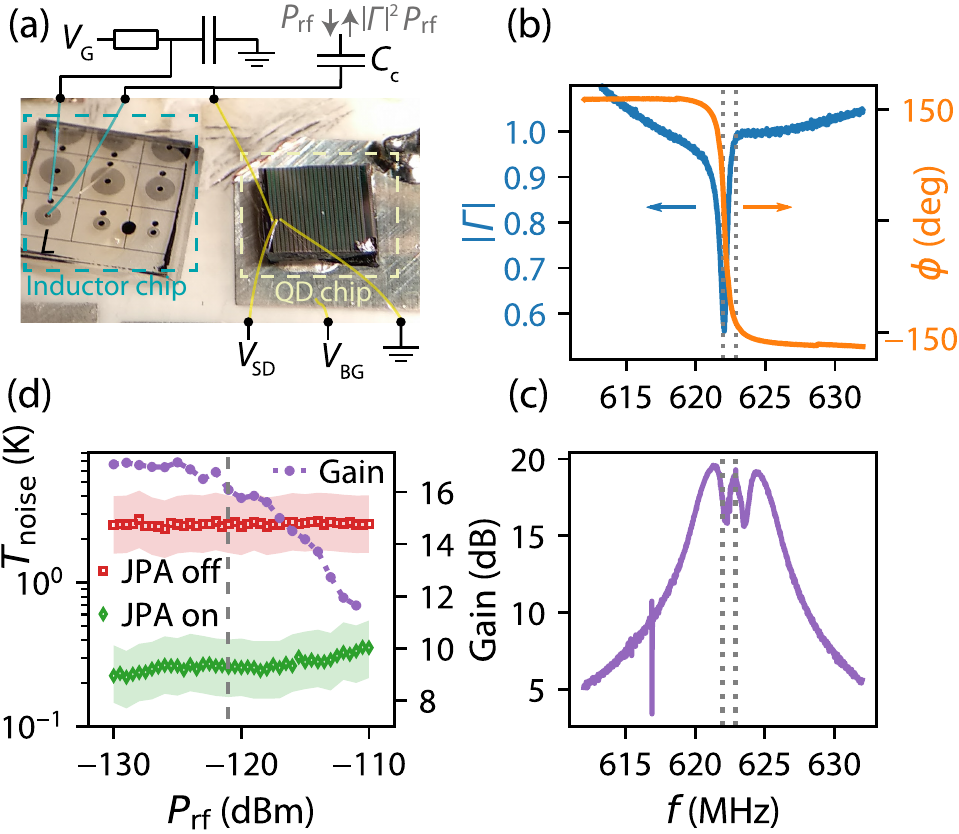}
		\caption{{\bf Characterizing the $LC$ resonator and its operation with the JPA}. \textbf{(a)} Shows the resonant circuit, including a schematic of the on-PCB components and bond-wires, the NbN spiral inductor, and the quantum dot device on a separate chip. \textbf{(b)} Magnitude and phase of the reflection coefficient showing the readout resonator ($Q_\mathrm{load}=966$). \textbf{(c)} Gain profile of the JPA when tuned close to the readout resonator frequency (3-dB-bandwidth of $6$~MHz). \textbf{(d)} JPA gain and estimated system noise temperature at $f_\mathrm{rf}$ as a function of RF input power showing saturation at high input power (1-dB-compression at $-121$~dBm).} 
		\label{fig:resonator}
	\end{figure}
	
	The dynamic range of this JPA is of the order of $-130$~dBm, making it unsuitable for the signal powers commonly used in previous reflectometry measurements ($-90$ to $-80$~dBm) ~\cite{Gonzalez-Zalba2015}. Here, we overcome this limitation using a high quality factor $LC$ resonator that is well coupled to the input line. The high $Q$-factor enables a reduced RF signal power to be used while achieving the same gate voltage on the device.
	The resonator circuit is formed by the parallel combination of a NbN spiral inductor $L=170$~nH, parasitic capacitance and the quantum dot device ($C_\mathrm{p}+C_\mathrm{d}=380$~fF), all coupled to the RF line via a coupling capacitor ($C_\mathrm{c}=37$~fF). We observe a resonance in the reflection coefficient $\Gamma = |\Gamma| \exp(i\phi)$ at $f_\mathrm{rf}=1/2\pi \sqrt{L (C_\mathrm{c} + C_\mathrm{p} + C_\mathrm{d})}=621.9$~MHz with a loaded quality factor of $Q_\mathrm{load}=966$, impedance $Z=\sqrt{L/(C_\mathrm{p}+C_\mathrm{c}+C_\mathrm{d})}=650\, \Omega$, return loss of 3~dB and phase shift $>180^\circ$ (over-coupled) as shown in Fig.~\ref{fig:resonator}(b). When operating at a charge instability in the QD device, the resonator reaches perfect matching.
	We operate the JPA in phase-preserving mode, where there is an offset $\Delta f=f_\mathrm{JPA}-f_\mathrm{rf}$ between the JPA pump frequency ($f_\mathrm{JPA}$) and $f_\mathrm{rf}$, so power from the JPA pump is transferred onto $f_\mathrm{rf}$ and $f_\mathrm{JPA}+\Delta f$ (four-wave mixing) via double-sideband phase modulation as illustrated by the transfer function. We select $\Delta f=1$~MHz to fall between the bandwidth of the resonator $\Delta f_\mathrm{rf}^\mathrm{3db} = 0.65$~MHz and the JPA $\Delta f_\mathrm{JPA}^\mathrm{3db} = 6$~MHz. This puts $f_\mathrm{JPA}$ at the edge of the readout resonator to avoid power broadening due to leakage of the pump signal while maximizing gain at $f_\mathrm{rf}$.
	When tuned and pumped, we achieve a gain of 17~dB at $f_\mathrm{rf}$ as shown in Fig.~\ref{fig:resonator}(c). The decrease in gain near $f_\mathrm{rf}$ is likely due to large impedance variations of the resonator close to resonance and imperfect matching to 50~$\Omega$.
	
	Figure~\ref{fig:resonator}(d) shows the JPA gain and the effective noise temperature close to $f_\mathrm{rf}$ as a function of $P_\mathrm{rf}$. We identify 1-dB-compression at $-121$~dBm. Based on amplifier gain estimations ($G_\mathrm{HEMT}=27\pm2$~dB) we obtain an effective noise temperature $T_\mathrm{noise}=2.5^{+1.4}_{-0.9}$~K with the JPA off (consistent with the cryogenic amplifier specifications) and a minimum noise temperature of $T_\mathrm{noise}=200^{+110}_{-73}$~mK based on the SNR improvement with the JPA on. 
	The effective noise temperature with the JPA on increases with increasing power due to saturation.
	There are multiple contributions to $T_\mathrm{noise}$, captured in Eq.~\ref{eq:Tsys}. We calculate the contribution of the cryogenic amplifier $\frac{T_\mathrm{HEMT}}{G_\mathrm{JPA}}=50^{+28}_{-18}$~mK and estimate $T_\mathrm{JPA}$ and $T_\mathrm{sys}$ by comparing $T_\mathrm{noise}$ when operating the QD device away from or at a charge instability. $T_\mathrm{sys}$ can have contributions from the resonator circuit ($T_\mathrm{circ}$) and the QD device ($T_\mathrm{QD}$): $T_\mathrm{sys}=(1-|\Gamma|^2) T_\mathrm{circ} + k T_\mathrm{QD}$~\cite{Muller2013}, where $k$ is the fraction of the rf power dissipated in the device. As we shall see later, in our device, tunneling between the QD and reservoir occurs adiabatically and hence $k=0$. Based on an increase in $T_\mathrm{noise}$ of $35^{+24}_{-13}$~mK when operating at a charge transition (where $|\Gamma|$ decreases from $0.5$ to $0$), we estimate $T_\mathrm{JPA}=47^{+35}_{-30}$~mK and $T_\mathrm{circ}=142^{+94}_{-54}$~mK. We relate $T_\mathrm{circ}$ to an electron temperature of dissipative elements in the resonant circuit and a we note that a JPA efficiency of 36\% of the quantum limit (equivalent to $\sim1.5$ photons) is compatible with previous results for operation close to a bifurcation point~\cite{Bryant1991,Hatridge2011,Boutin2017}. 
	
\begin{figure}
		\centering
		\includegraphics[width=\linewidth]{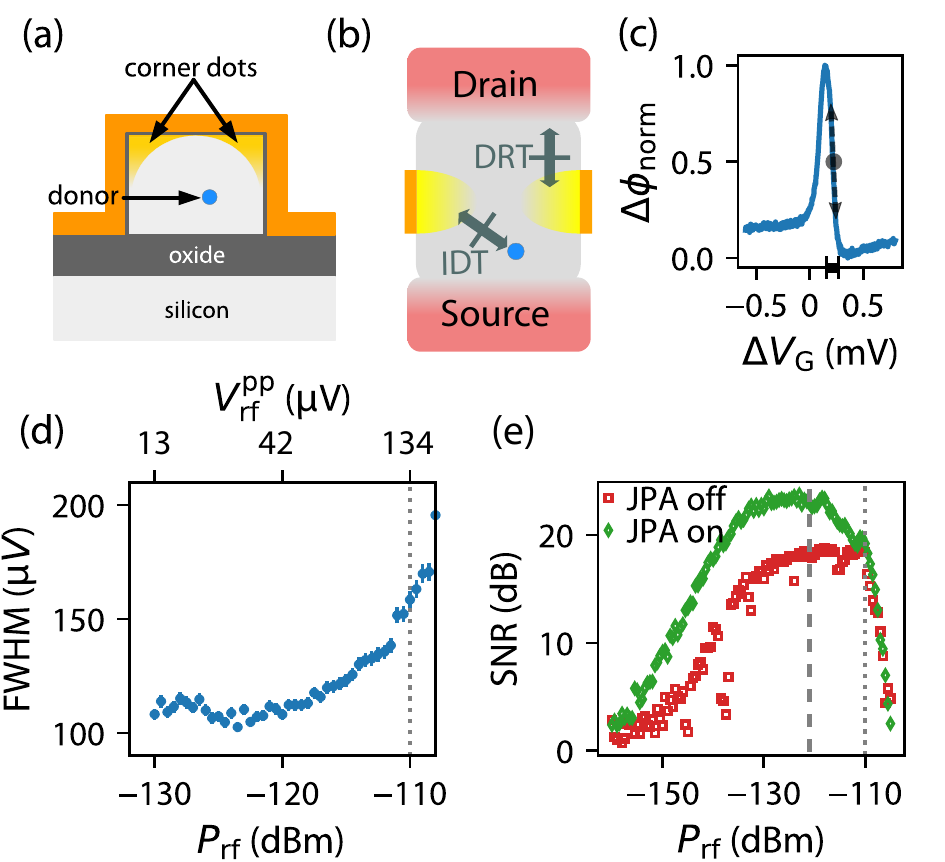}
		\caption{\textbf{Dot-to-reservoir transition (DRT) charge sensitivity.} \textbf{(a)} Schematic cross-section of the QD device along the gate, and \textbf{(b)} along the source and drain, showing two QDs in the top corners of the nanowire and a donor in the channel. Transitions at which measurements are performed are indicated. \textbf{(c)} Typical phase response across a DRT ($P_\mathrm{rf} = -125$~dBm). 
			\textbf{(d)} Full-width half maximum (FWHM) of the DRT transition as a function of power showing significant power broadening for $P_\mathrm{rf} > 110$~dBm. \textbf{(e)} SNR as a function of RF power of a charge sensitivity measurement at the DRT, indicating the power at which JPA saturates (dashed line) and significant power broadening occurs (dotted line).}
		\label{fig:qd}
	\end{figure}

	Next, we characterize and compare the improvements in the SNR of gate-based readout using a quantum dot-to-reservoir transition (DRT) and inter-dopant/dot charge transition (IDT) in a CMOS silicon nanowire field-effect transistor device with channel width and gate length of 30~nm and nanowire height of 11~nm~\cite{Urdampilleta2015}. Figure~\ref{fig:qd}(a) shows a schematic line-cut of the device along the gate (see SEM micro-graph in Fig.~\ref{fig:jpa}(a)). QDs form in the corners of the device and have a strong coupling to the gate $\alpha_{\rm DRT}=0.86$. Additionally, given the doping density, an average of 5  (phosphorus) donors are expected in the device channel~\cite{Urdampilleta2015}. 
	%
	We first focus on a DRT transition in the device, in which a QD is primarily tunnel coupled to the drain reservoir as illustrated in a schematic line-cut of the device along the source-drain direction in Fig.~\ref{fig:qd}(b). When operating at the DRT transition we observe a capacitive shift of the resonance corresponding to $\Delta C_\mathrm{d}=0.5$~fF. In Fig.~\ref{fig:qd}(d) we show the full-width half maximum of the selected transition as a function of $P_\mathrm{rf}$. No power broadening is observed for $P_\mathrm{rf}$ below $-120$~dBm, while the transition is significantly broadened above $-110$~dBm. Due to the high $Q_\mathrm{load}$ of the resonator, only a small input power, compatible with the dynamic range and saturation of the JPA, is required and we calculate the RF disturbance at the device gate as $V_\mathrm{rf}^\mathrm{pp}=\frac{2 C_\mathrm{c}}{C_\mathrm{c} + C_\mathrm{p} + C_\mathrm{d}} Q_\mathrm{load} V_\mathrm{in}^\mathrm{pp}$, corresponding, for example, to  $V_\mathrm{rf}^\mathrm{pp}=13~\upmu$V for an input RF power of $-130$~dBm. 
	
	Next, we use conventional methods to measure the charge sensitivity~\cite{SuppMat,Schoelkopf1998} with and without the JPA, which provides a device-specific benchmark on the performance of our gate-based sensor normalized to the gate charge. The SNR as a function of $P_\mathrm{rf}$, with and without the JPA, is shown in Fig.~\ref{fig:qd}(e), when operating at a small gate voltage modulation of 50~kHz as indicated in Fig.~\ref{fig:qd}(c). We observe an improvement of up to 8~dB in SNR with the JPA at low RF power. Irrespective of whether the JPA is used, for $P_\mathrm{rf}$ between $-130$ and $-120$~dBm the SNR levels off as DRT begins to become power broadened, and it drops abruptly for powers above $-110$~dBm. With the JPA on there is an additional decrease in SNR evident above $-120$~dBm as the JPA saturates. The JPA can either be used to increase the SNR beyond what could otherwise be achieved, and/or to provide the same SNR but at about 10~dB less RF power, with the corresponding reduction in the disturbance of the QD being measured, and its neighbors.
	When operating well below power broadening ($P_\mathrm{rf}=-130$~dBm), the charge sensitivity achieved with the JPA is $0.25\, \upmu e/\sqrt{\mathrm{Hz}}$ compared to $0.5\, \upmu e/\sqrt{\mathrm{Hz}}$ without the JPA, outperforming previous measurements using RF-SET~\cite{Brenning2006} and gate-based approaches~\cite{Ahmed2018}.
	%
	\begin{figure}
		\centering
		\includegraphics[width=\linewidth]{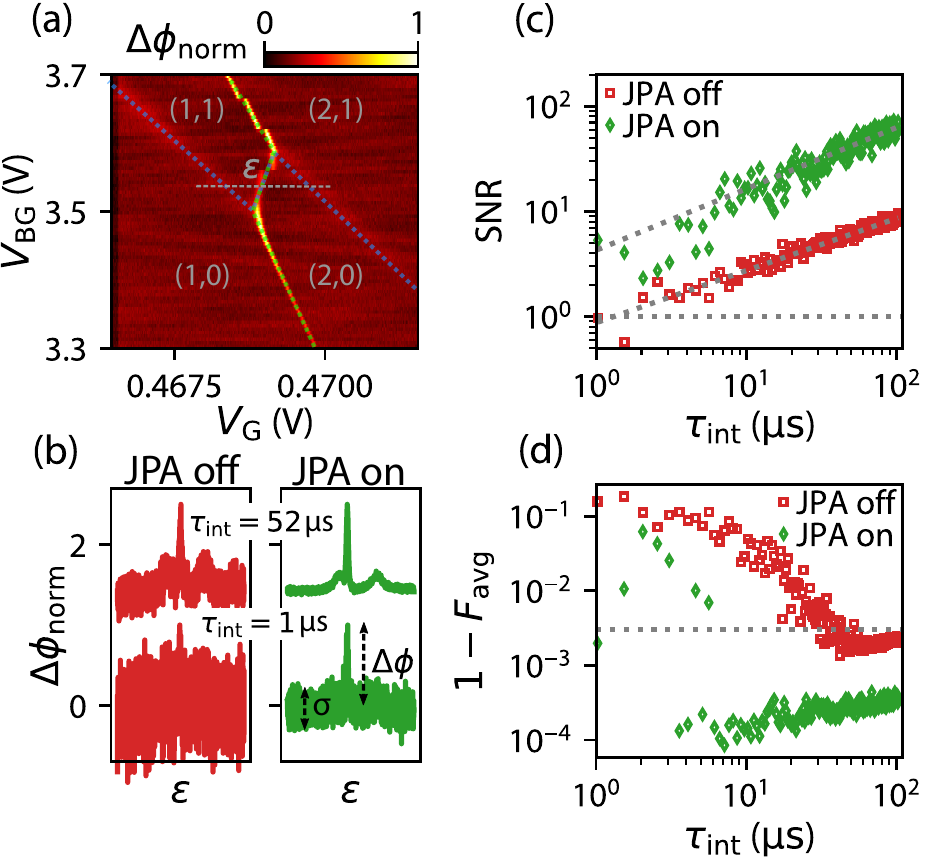}
		\caption{\textbf{Inter-donor/dot charge transition (IDT). }\textbf{(a)} Even-parity IDT between a donor and QD in the device. Effective electron occupation ($N_\mathrm{dot}$, $N_\mathrm{donor}$) indicated, up to an arbitrary offset. \textbf{(b)} Normalized phase response along the detuning axis $\varepsilon$ in (a) with and without the JPA for two different integration times (traces offset by 1.5 in $\Delta \phi$ for clarity). \textbf{(c)}~SNR obtained from traces as shown in (b) as a function of integration time with JPA on and off. Linear extrapolation and $\text{SNR}=1$ is indicated using dotted lines. \textbf{(d)} Average readout infidelity $1-F_\mathrm{avg}$ as a function of $\tau_\mathrm{int}$ obtained from simulations of the singlet and triplet probability distributions based on the signal and noise levels shown in (c) for the JPA off and on. The horizontal dotted line indicates $F_\mathrm{avg}=0.997$. $F_\mathrm{avg}>0.997$ is obtained at $\tau_\mathrm{int}=32\, \upmu$s with the JPA off and $\tau_\mathrm{int}=1\, \upmu$s with the JPA on. }
		\label{fig:dopant}
	\end{figure}
	
	Finally, we move on to benchmark the readout using a donor-dot IDT of even parity~\cite{SuppMat} that exhibits features of spin blockade~\cite{Schroer2012a}. 
	Figure~\ref{fig:dopant}(a) shows the IDT in the normalized phase response as a function of $V_\mathrm{G}$ and back-gate voltage ($V_\mathrm{BG}$), where the donor transition can be identified due to a stronger coupling to $V_\mathrm{BG}$ (donor resides deeper in the channel, closer to the back-gate). We determined the charge occupation of the donor and QD (up to an arbitrary offset) using the magnetic field response and we calculate a gate coupling $\alpha_{\rm IDT}=0.36$ 
	and tunnel coupling $\Delta_\mathrm{c}=20.9\, \upmu$eV, corresponding to a capacitive shift of $\Delta C_\mathrm{d}=0.5$~fF~\cite{SuppMat}.
	Figure~\ref{fig:dopant}(b) shows an example of the phase response across the IDT, along the detuning axis $\varepsilon$ shown in Fig.~\ref{fig:dopant}(a) for $\tau_\mathrm{int}=1\, \upmu$s and $\tau_\mathrm{int}=52\, \upmu$s.
	We determine the power SNR from the amplitude $\Delta \phi$ of the IDT signal and the RMS amplitude $\sigma$ of the noise as $\text{SNR}=\frac{\Delta \phi^2}{\sigma^2}$.
	Figure~\ref{fig:dopant}(c) shows the SNR as a function of integration time with the JPA on and off. We observe an improvement in SNR of a factor of 7 with the JPA on consistent with the SNR improvement of 8~dB observed in the charge sensitivity measurement. Using an extrapolation (dotted lines in Fig.~\ref{fig:dopant}(c)) we infer an SNR of unity at $\tau_\mathrm{int}^\mathrm{off}=1.2\, \upmu$s and $\tau_\mathrm{int}^\mathrm{on}=80\, $ns with the JPA off and on, respectively. However, the limited bandwidth of our resonator prohibits measurements faster than $1.5\, \upmu$s ($\Delta f^\mathrm{3db}_\mathrm{rf}=0.65$~MHz). Additionally, we observe that multiple measurements of the SNR with the JPA on for $\tau_\mathrm{int}^\mathrm{on}<10\, \upmu$s deviate from the extrapolation, which could be due to noise introduced by the JPA pump signal which is operated only 1~MHz offset the RF signal.
	Based on the signal and noise levels (as shown in Fig.~\ref{fig:dopant}(c)) we simulate the singlet and triplet readout probability densities~\cite{SuppMat}, modeled as two noise-broadened Gaussian distributions~\cite{Barthel2009}. The model includes terms to account for relaxation of the triplet during measurement and we assume $T_1=4.5$~ms~\cite{West2019}. From the probability densities we obtain an average readout infidelity as a function of integration time which is shown Fig.~\ref{fig:dopant}(d) for the JPA off and on. We find that $F_\mathrm{avg}>0.997$ can be reached at a $\text{SNR}>5$ corresponding to an integration time of at least $\tau_\mathrm{int}^\mathrm{off}=32\, \upmu$s with the JPA off and $\tau_\mathrm{int}^\mathrm{on}=1\, \upmu$s with the JPA on allowing readout faster than the the coherence time of electron spins in $^{28}$Si ($T_2^*= 120$~$\upmu$s~\cite{Veldhorst2014}).
	
	
	We have demonstrated that the SNR of RF gate-based readout of quantum dot devices can be improved using a JPA. We observe a SNR improvement of 8~dB for both dot-to-reservoir and inter-donor/dot transitions when the JPA is operated closed to the RF frequency in phase-preserving mode at 17~dB gain.
	We have analyzed the performance of the JPA in an external magnetic field, commonly applied in spin qubit devices, and find no disturbance on the JPA performance up to a field of $B_z=3$~T at the device~\cite{SuppMat}.  
	The SNR improvement we see is currently limited by the gain of the JPA, yielding a contribution of the cryogenic amplifier of at least 50~mK to the system noise temperature. Assuming a JPA gain of 23~dB or more, the contribution of the cryogenic amplifier would become negligible. The system noise performance could be further improved by operating the JPA in phase-sensitive mode, where the noise added by the JPA can be squeezed below the quantum limit. Changes in the circuit such as additional isolators between the JPA and readout resonator as well as additional line attenuation and filtering could be beneficial towards achieving larger gain, a lower system noise temperatures and prevent leakage of the JPA pump signal into the readout resonator.
	In addition, the measurement speed in this implementation is, in principle, limited by the bandwidth of our high-$Q$ readout resonator: increasing the coupling to the line or, preferentially, moving to a higher frequency of the resonator circuit while maintaining high loaded $Q$ should allow sub-microsecond fault-tolerant gate-based spin readout. 
	Further development could reduce the footprint of the high $Q$ resonators, to achieve an integrated and scalable readout architecture~\cite{Schaal2018a} with the potential of reduced circuit losses and parasitics. Using a traveling wave parametric amplifier (TWPA) with increased bandwidth, frequency multiplexing of multiple resonators could be achieved.

	\begin{acknowledgments}
		This research has received funding from the European Union's Horizon 2020 research and innovation programme under grant agreement No 688539 (http://mos-quito.eu) and Seventh Framework Programme (FP7/2007-2013) through Grant Agreement No. 318397 (http://www.tolop.eu.); as well as by the Engineering and Physical Sciences Research Council (EPSRC) through the Centre for Doctoral Training in Delivering Quantum Technologies (EP/L015242/1), UNDEDD (EP/K025945/1) and QUES2T (EP/N015118/1). J.W.A.R. acknowledges funding from the EPSRC through International Network and Programme Grants (EP/P026311/1; EP/N017242/1). JPA development was supported by the Army Research Office under Grant No. W911NF-14-1-0078. M.F.G.Z. acknowledge support from the Royal Society and Winton Programme for the Physics of Sustainability.
	\end{acknowledgments}

\end{document}


	
	

\title{Supplemental Material: Fast gate-based readout of silicon quantum dots using Josephson parametric amplification}
	
	\author{S. Schaal}
\email{simon.schaal.15@ucl.ac.uk}
\affiliation{London Centre for Nanotechnology, UCL, London WC1H 0AH, United Kingdom}

\author{I. Ahmed}
\altaffiliation[Present address: ]
{Dept.\ of Electrical \& Electronic Engineering, Univ.\ of Dhaka, Dhaka 1000, Bangladesh}
\affiliation{Cavendish Laboratory, Univ.\ of Cambridge, J. J. Thomson Ave., Cambridge, CB3 0HE, U.K.}
\author{J. A. Haigh}
\affiliation{Hitachi Cambridge Laboratory, J.J. Thomson Avenue, Cambridge CB3 0HE, United Kingdom}

\author{L. Hutin}
\author{B. Bertrand}
\author{S. Barraud}
\author{M. Vinet}
\affiliation{CEA, LETI, Minatec Campus, F-38054 Grenoble, France}

\author{C.--M. Lee}
\author{N. Stelmashenko}
\author{J. W. A. Robinson}
\affiliation{Dept.\ of Materials Science \& Metallurgy, Univ.\ of Cambridge, 27 Charles Babbage Road, CB3 0FS, U.K.}

\author{J. Y. Qiu}
\altaffiliation[Present address: ]
{Massachusetts Institute of Technology, Cambridge, MA 02139, USA}
\affiliation{Quantum Nanoelectronics Laboratory, Dept.\ of Physics, 	Univ.\ of California, Berkeley CA 94720, USA}	
\author{S. Hacohen-Gourgy}
\author{I. Siddiqi}
\affiliation{Quantum Nanoelectronics Laboratory, Dept.\ of Physics, 	Univ.\ of California, Berkeley CA 94720, USA}	

\author{M. F. Gonzalez-Zalba}
\affiliation
{Hitachi Cambridge Laboratory, J.J. Thomson Avenue, Cambridge CB3 0HE, United Kingdom}

\author{J. J. L. Morton}
\email{jjl.morton@ucl.ac.uk}
\affiliation{London Centre for Nanotechnology, UCL, London WC1H 0AH, United Kingdom}
\affiliation
{Dept.\ of Electronic \& Electrical Engineering, UCL, London WC1E 7JE, United Kingdom}

\maketitle



	\section{\label{sec:setup}Measurement setup and fabrication details}
	\begin{figure}[H]
		\centering
		\includegraphics[width=0.7\linewidth]{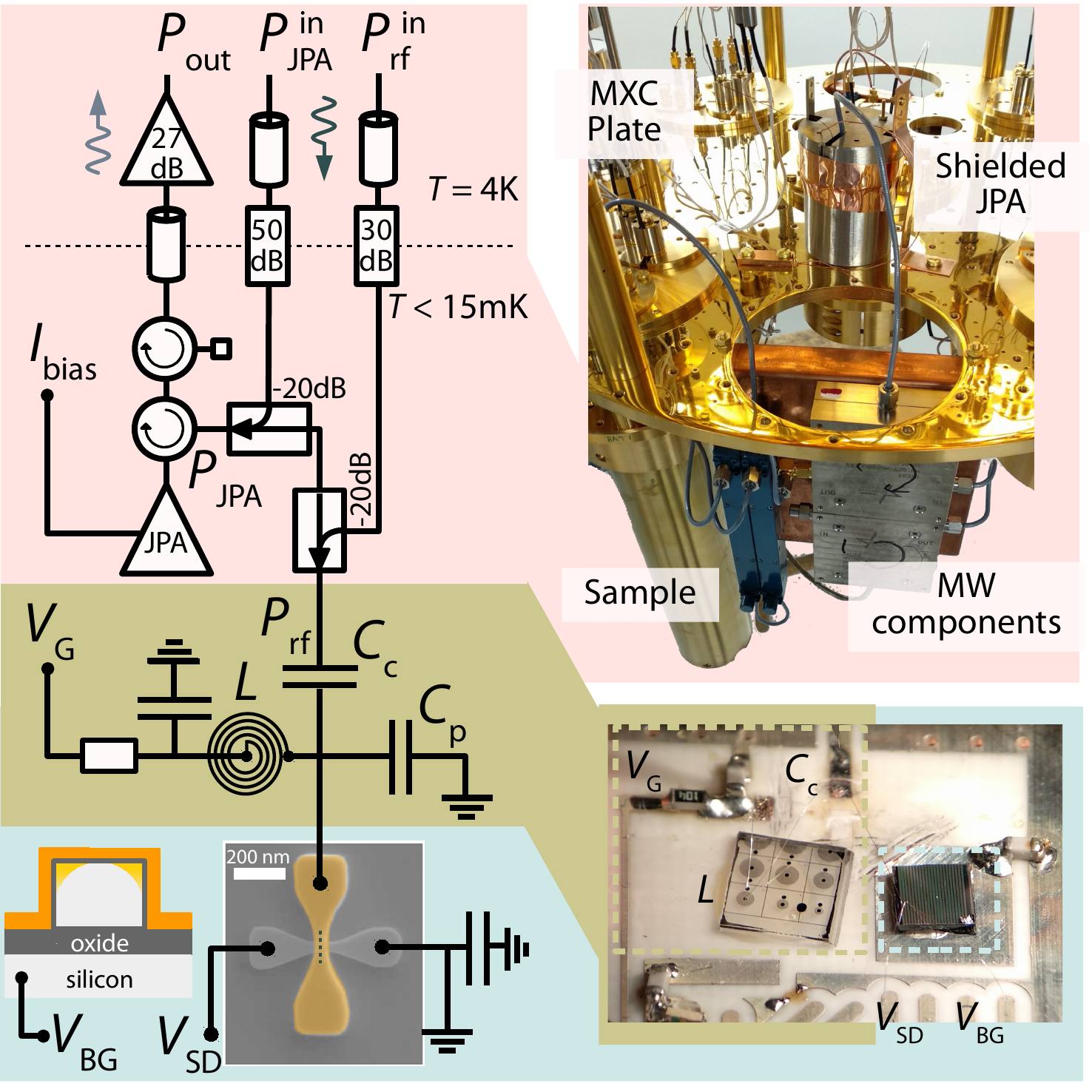}
		\caption{Experimental setup with pictures in a fridge without magnet.}
		\label{fig:setup}
	\end{figure}

	Measurements were performed at base temperature of a dilution refrigerator ($10$~mK). The cryogenic parts of the experimental setup with pictures is shown in Fig.~\ref{fig:setup}. Low frequency signals ($V_{\mathrm{SD}}$, $V_{\mathrm{G}}$, $V_{\mathrm{BG}}$) were delivered through filtered cryogenic loom (combination of RF and RC filter) while a radio-frequency signal for gate-based readout was delivered through an attenuated and filtered coaxial line via a cryogenic directional coupler (Krytar, indicated attenuation includes contribution from the coupler) which connects to a on-PCB bias tee combining the RF  modulation with $V_{\mathrm{G}}$. The PCB is made from $0.5$~mm thick RO4003C with immersion silver finish.
	The resonator is formed from a $170$~nH NbN spiral inductor, a coupling capacitor and the sample's parasitic capacitance to ground in parallel with the device. The NbN spiral inductor is fabricated on a $430\, \mu$m thick c-plane sapphire substrate. A 80~nm thick NbN films was grown using unheated DC magnetron sputtering in an Ar/N2 atmosphere with $28\%$ N2 at 1.5~Pa. The spiral with $8\, \mu$m feature size was etched by means of optical lithography and is wire-bonded to the gate of the nanowire-transistor device.
	The reflected RF  signal passes the Josephson parametric amplifier (JPA) via cryogenic couplers and circulators (Quinstar QCY) which are thermally anchored to the mixing chamber plate (MXC) followed by additional amplification at $4$~K (LNF-LNC0.6\_2A) and room temperature (ZRL-700+) and final quadrature demodulation (Polyphase Microwave AD0540B) from which the amplitude and phase of the reflected signal is obtained (homodyne detection).
	The JPA and circulators are fitted with cryoperm shields that withstand fields up to 1500~G. The maximum stray field at the MXC (40~cm from field center) at $B_z=6$~T is 500~G. No field compensation was used.
	
	Nanowire transistors devices used in this study were fabricated on SOI substrates with a $145$-nm-thick buried oxide and $10$-nm-thick silicon layer. The silicon layer is patterned to create the channel by means of optical lithography, followed by a resist trimming process. All transistors share the same gate stack consisting of $1.9$~nm HfSiON capped by $5$~nm TiN and $50$~nm polycrystalline silicon leading to a total equivalent oxide thickness of $1.3$~nm. The Si thickness under the HfSiON/TiN gate is $11$~nm. After gate etching a SiN layer ($10$~nm) was deposited and etched to form a first spacer on the sidewalls of the gate. 18-nm-thick Si raised source and drain contacts were selectively grown before the source/drain extension implantation and activation annealing. A second spacer was formed followed by source/drain implantations, activation spike anneal and salicidation (NiPtSi).

	\section{\label{sec:newdev}Donor-dot transition}	
	
\begin{figure}
		\centering
		\includegraphics[width=0.7\linewidth]{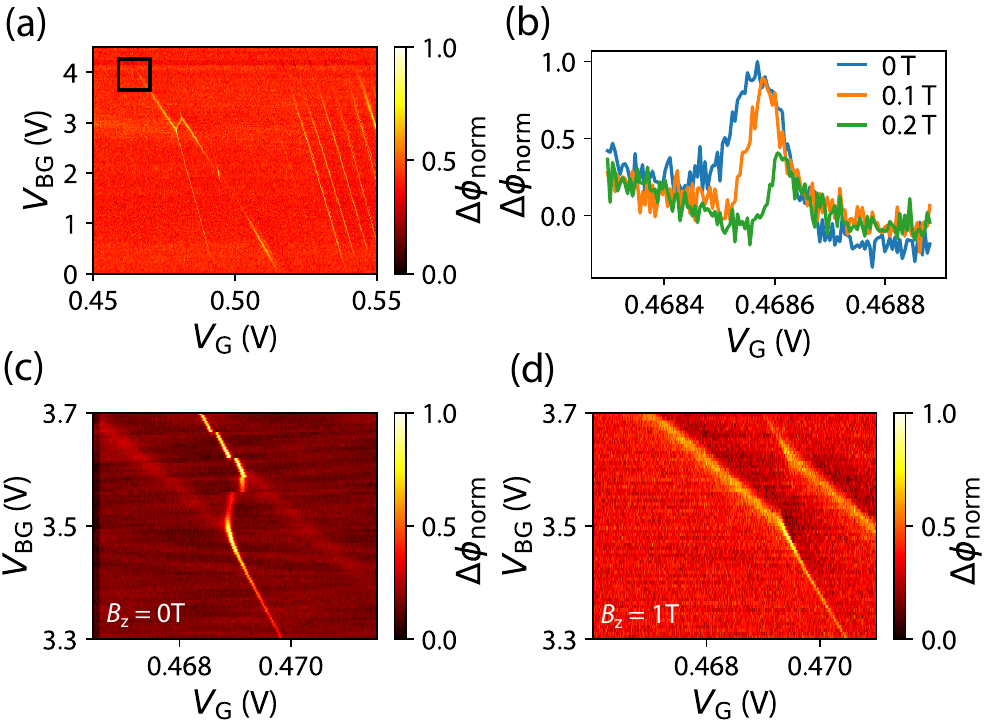}
		\caption{Dopant-dot device. \textbf{(a)} Several dot-to-reservoir transitions (DRT) and one dopant transition is observed in a second device. Relevant inter-dopant/dot charge transition (IDT) highlighted at the top. \textbf{(b)} Line scan of the IDT for at different external field. Signal slowly disappears with increasing field indicating even charge parity. \textbf{(c-d)} IDT shown for $B_z=0$~T and $B_z=1$~T respectively. At 1~T the inter-dot signal disappears demonstrating spin-related effects of even parity for the IDT.}
		\label{fig:dotdopant}
	\end{figure}

	An exemplary measurement of the phase response of a quantum device presented in the main text is shown Fig.~\ref{fig:dotdopant}(a) as a function of top-gate and back-gate voltage. Multiple nearly parallel lines can be observed which can be attributed to dot-to-reservoir transitions of quantum dots in the device which couple strongly to the top-gate. One additional line is apparent which couples more strongly to the back-gate and anti-crosses with three of the quantum dot transitions. We attribute this transition to a phosphorus donor. The inter-dopant/dot charge transition (IDT) discussed in the main text is highlighted at the top left.
	From a reduction of the IDT signal with increasing magnetic field as shown in Fig.~\ref{fig:dotdopant}(b-d) we conclude that the IDT is of even parity where singlet and triplet states are formed. As the magnetic field is increased the triplet $T_-$ state becomes the ground state. Due to the Pauli exclusion principle the triplet energy is linear as a function of detuning which results in zero charge susceptibility compared to the singlet which has a large curvature at zero detuning. In Fig.~\ref{fig:dotdopant}(b) the IDT shifts towards higher top-gate voltage which indicates that the (2,0) charge configuration is found at positive detuning where the $T_-$(1,1) and $S$(2,0) states cross. In this analysis we have assigned a (1,1)-(2,0) charge occupation to the IDT based on the fact that it is an even parity transition and we couldn't observe additional lines at lower voltage. However, there could be an arbitrary offset in the number of electrons. Additionally, from the shift to larger voltage with increasing field and the expected lineshape of the capacitive signal transition we obtain the gate coupling $\alpha=0.36$~eV/V, tunnel coupling $\Delta_\mathrm{c}=20.9\, \mu$eV and maximum capacitive shift $C_0=0.5$~fF
	
	\begin{align}
		\alpha V_\mathrm{G} &= \frac{\Delta_\mathrm{c}^2 - (2 g \mu_B B)^2}{4g\mu_B B} \approx -g\mu_B B, \quad B\gg \Delta_\mathrm{c}\\
		\Delta C_\mathrm{d} &= C_0 \frac{\Delta_\mathrm{c}^3}{((\alpha V_\mathrm{G})^2 + \Delta_\mathrm{c}^2)^{3/2}} \tanh\left({\frac{\sqrt{(\alpha V_\mathrm{G})^2 + \Delta_\mathrm{c}^2}}{2k_B T}}\right) \\
		C_0 &= \frac{(\alpha e)^2}{2 \Delta_\mathrm{c}}.
	\end{align}

	\section{\label{sec:sens}Charge sensitivity}
\begin{figure}
	\centering
	\includegraphics[width=0.7\linewidth]{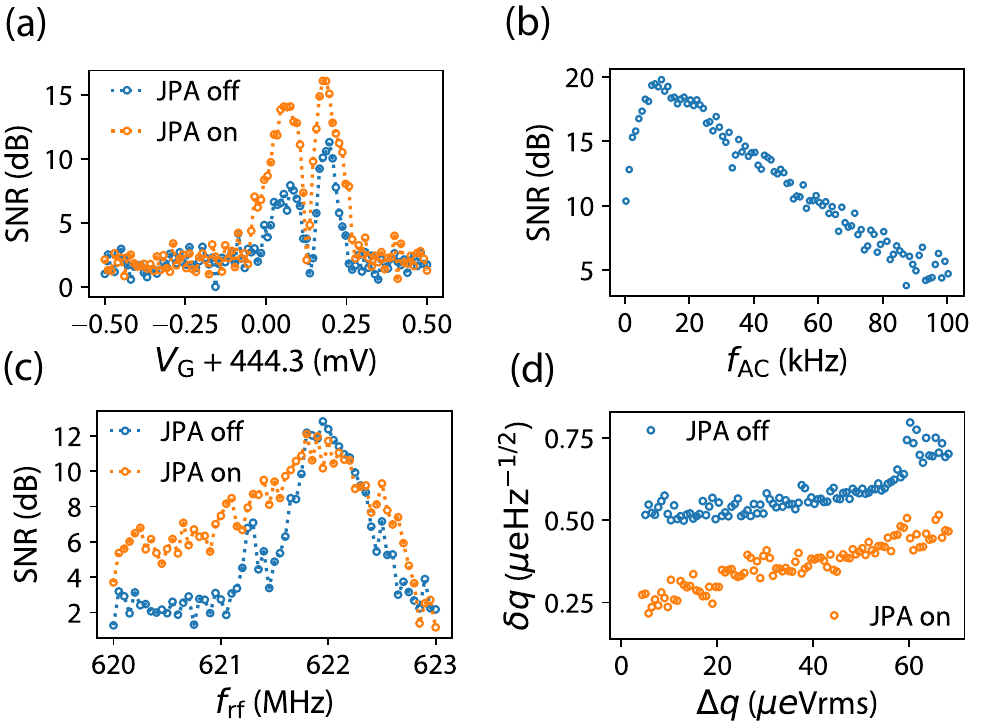}
	\caption{Fine tuning for charge sensitivity measurements. \textbf{(a)} Fine tuning of the gate voltage offset. Highest SNR is observed at steepest slope of dot-to-reservoir transition. \textbf{(b)} Signal-to-noise as a function of modulation frequency at $P_\mathrm{rf}=-120$~dBm. Fine tuning of RF frequency \textbf{(c)} and modulation amplitude \textbf{(d)} for sensitivity measurements.}
	\label{fig:sens}
\end{figure}

For measurements of the charge sensitivity a small sinusoidal modulation is applied to the top-gate of the device at frequency $f_\mathrm{AC}$ and root-mean-square charge equivalent amplitude $\Delta q$.
In order to find the optimal charge sensitivity multiple parameters need to be tuned. First, the gate-voltage offset needs to be set to the point of the largest slope of the DRT transition where the maximum SNR is observed in Fig.~\ref{fig:sens}(a). Next the modulation frequency can be optimized. In this particular device we observe that initially the SNR increases with $f_\mathrm{AC}$ in Fig.~\ref{fig:sens}(b) and has a maximum at about 10~kHz followed by a decrease at higher frequency. We attribute the low SNR close to the RF carrier to charge noise in the device which increases with power and is insignificant when operating at small RF power below power broadening $P_\mathrm{rf}<-130$~dBm. At $f_\mathrm{AC}>10$~kHz the SNR reduces due to a amplitude reduction of the modulation due to filters in the line. We choose to operate at 50~kHz to minimize the effect of charge noise on the charge sensitivity measurement and calibrate the amplitude reduction due to filters on the line at 50~kHz to 11\%.
Figure~\ref{fig:sens}(c) shows the SNR as a function of $f_\mathrm{rf}$ with the JPA off (at $P_\mathrm{rf}=-130$~dBm) and on (at $P_\mathrm{rf}=-120$~dBm), showing the optimal operation frequency.
Finally, Fig.~\ref{fig:sens}(d) shows the charge sensitivity, calculated as $\delta q = \Delta q / (\sqrt{2 \cdot BW} \times 10^{\frac{\text{SNR}}{20}})$, as a function of modulation amplitude ((at $P_\mathrm{rf}-130$~dBm) which shows that best sensitivity is achieved at small modulation.  

	\section{\label{sec:magn}Magnetic field}

\begin{figure}
	\centering
	\includegraphics[width=0.7\linewidth]{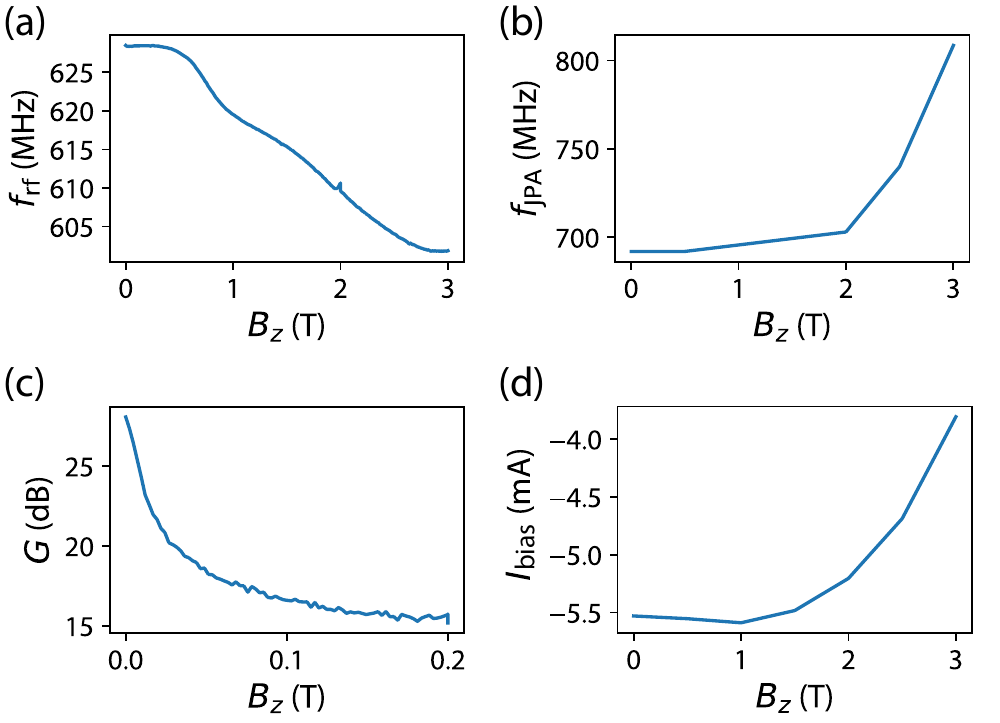}
	\caption{Magnetic field. \textbf{(a)} Resonant frequency of the NbN based readout resonator as a function of external magnetic field $B_z$. \textbf{(b)} JPA frequency as a function of $B_z$. \textbf{(c)} Gain of the JPA tuned to 28~dB gain at 644~MHz and $B_z=0$~T as a function of external field. \textbf{(d)} Flux bias $I_\mathrm{bias}$ as a function of $B_z$ required to recover 28~dB gain at 644~MHz.}
	\label{fig:field}
\end{figure}

Spin qubits in semiconductor devices are usually operated at fixed finite magnetic fields. Both the readout resonator and the JPA are superconducting devices which are affected by a magnetic field. 
Here, we investigate the effect of an external magnetic field on both the resonator and JPA where the field $B_z$ is the field at the sample box in the center of the magnet. The resonator resides in the sample box while the JPA is mounted on the mixing chamber plate 40~cm away from the field center where a maximum of 500~G is expected at $B_z=6$~T.
The field $B_z$ leads to a reduction of the readout resonator frequency as shown in Fig.~\ref{fig:field}(a) due to kinetic inductance. At first, up to 0.5~T, the resonance does not change significantly and then reduces by about 30~MHz from 0.5~T to 3~T. Similarly, the JPA resonance changes with applied field. The JPA is especially susceptible to external fields as its resonance is tuned using the tunable inductance of a SQUID loop. Any field penetrating into the loop tunes the frequency. Additionally, any field noise strongly affects the JPA performance. As a result of this, the JPA is kept in cryoperm magnetic shield as shown in Fig.~\ref{fig:setup}. In Fig.~\ref{fig:field}(b) we observe that the JPA resonance merely changes up to 2~T after which some of the external field penetrates into the shield resulting in an increased frequency of about 100~MHz at 3~T.
Next, we examine how the gain of 28~dB of the JPA operating at $644$~MHz is affected by a magnetic field. In Fig.~\ref{fig:field}(c) we see that the gain drops quickly and settles towards 15~dB at 0.2~T. 

While this is a significant change in gain, we find that the gain of 28~dB can be restored by tuning a single parameter only, namely the flux bias $I_\mathrm{bias}$ as shown in Fig.~\ref{fig:field}(d) which shows the flux bias required to restore 28~dB of gain at $644$~MHz for a given magnetic field. This illustrates that up to 3~T the performance of the JPA is not reduced and restoring the gain after a change in magnetic field could be easily automated by optimizing a single parameter. However, at some point additional tuning might be necessary once the readout resonator frequency has shifted beyond the bandwidth of the JPA due to a change in external field.

	\section{Readout fidelity model}
Here, we give details to a model used in the main text to estimate the readout fidelity based on the signal and noise observed in the experiment. 
The signal we consider is the difference in the reflected phase of the RF signal between a singlet and triplet state $\Delta \phi = \phi_S-\phi_T$ of an even parity charge transition which arises from the difference in charge susceptibility due to the Pauli exclusion principle.
We model the singlet and triplet probability density as two Gaussian distributions separated by $\Delta\phi$ and broadened by the measurement noise $\sigma$. The outcome of $N_\mathrm{total}$ single shot experiments should then follow a histogram given by 
$$N(\Delta\phi)=N_\mathrm{total} ( (1-p_T) P_S(\Delta\phi) + p_t P_T(\Delta\phi) ) \Delta \phi_\mathrm{binsize}$$
where $p_T$ is the triplet probability over all outcomes, which we select as $p_T=0.5$, and $\phi_\mathrm{binsize}$ is the histogram bin size. We model the singlet probability density as 
$$P_S(\Delta\phi)=\frac{1}{\sqrt{2\pi} \sigma} \exp\left( \frac{-(\Delta\phi - \phi_S)^2}{2\sigma^2} \right),$$
while the triplet probability includes additional terms accounting for the relaxation of the triplet during the integration time $\tau_\mathrm{int}$ given the relaxation time $T_1$ of the blockade

\begin{figure}
	\centering
	\includegraphics[width=0.6\linewidth]{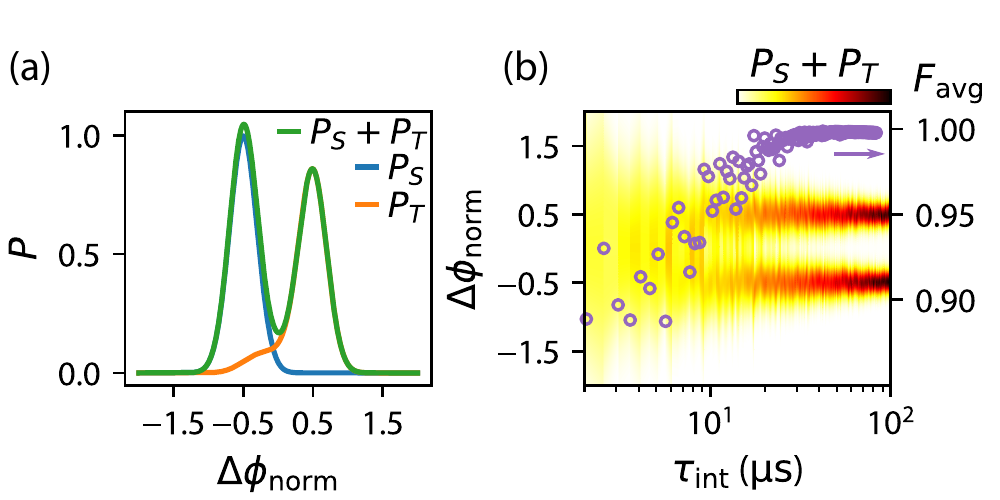}
	\caption{Readout fidelity model. \textbf{(a)} Exemplary probability distribution $P_{S/T}$ for $\sigma=0.2\cdot \Delta\phi_\mathrm{norm}$ and $\tau_\mathrm{int}=0.2 \cdot T_1$. Due to relaxation of the triplet to a singlet within the integration time, $P_T$ shows a significant probability at $\Delta\phi_\mathrm{norm}=-0.5$. \textbf{(b)} Shows the singlet and triplet probability distribution $P_{S} + P_T$ as a function of integration time for the signal and noise measured without the JPA and presented in Fig.~4(c) of the main text. On the right axis the average fidelity $F_\mathrm{avg}$ derived from the probability distributions is shown.} 
	\label{fig:fid}
\end{figure}

\begin{align*}
P_T(\Delta\phi) &= \frac{1}{\sqrt{2\pi} \sigma} \Bigl[ e^{-\tau_\mathrm{int}/T_1}  \exp\left( \frac{-(\Delta\phi - \phi_T)^2}{2\sigma^2} \right) \\
&+ \int_{\phi_S}^{\phi_T} \frac{\tau_\mathrm{int}}{T_1 (\phi_S-\phi_T)}  \exp\left(-\frac{\tau_\mathrm{int}}{T_1} \frac{\phi - \phi_S}{\phi_S-\phi_T}\right) \exp\left( \frac{-(\Delta\phi - \phi)^2}{2\sigma^2} \right) d\phi\ \Bigr].
\end{align*}

From the probability density the singlet and triplet fidelity is given by
\begin{align*}
	F_S&=1-\int_{\phi_\mathrm{th}}^{\infty} P_S(\Delta\phi) d\Delta\phi & F_T&=1-\int_{-\infty}^{\phi_\mathrm{th}} P_T(\Delta\phi) d\Delta\phi 
\end{align*}
where $\phi_\mathrm{th}$ is the selected phase threshold and the average fidelity and visibility is defined as
\begin{align*}
F_\mathrm{avg}&=\frac{F_S}{2} + \frac{F_T}{2} & V&=F_S + F_T -1.
\end{align*}

In our analysis we normalize the signal $\Delta\phi$ (and noise $\sigma$) such that we expect two Gaussian's separated by $\Delta\phi_\mathrm{norm}=1$. Figure~\ref{fig:fid}(a) shows exemplary singlet and triplet distributions with an integration time $\tau_\mathrm{int}$ comparable to $T_1$. In this regime a threshold $\phi_\mathrm{th}$ which is not perfectly centered between $P_S$ and $P_T$ yields the optimal fidelity. The probability distributions $P_S + P_T$ produced from the signal and noise obtained in the main text and the average fidelity as a function of $\tau_\mathrm{int}$ is shown in Fig.~\ref{fig:fid}(b).
When calculating the (in)fidelity we optimize $\phi_\mathrm{th}$ for every $\tau_\mathrm{int}$ and obtain $F_\mathrm{avg}=0.9978$ ($V=0.9954$) with $\phi_\mathrm{th}=-0.025$ at $\tau_\mathrm{int}=32\, \mu$s ($\mathrm{SNR}>5$) without the JPA and obtain the same fidelity with the JPA on at $\tau_\mathrm{int}=1\, \mu$s.
